\documentclass[prb,aps,twocolumn]{revtex4-2}

\usepackage{graphicx}
\usepackage{dcolumn}
\usepackage{amsmath,amssymb,graphicx,bm,epsfig}
\usepackage{epstopdf}
\usepackage{color}
\usepackage{tabularx}

\renewcommand{\a}{\alpha}

\begin{document}
\def\EF{$E_\textrm{F}$}
\def\ZF{$Z_\textrm{F}$}
\def\ZT{$\mathbb{Z}_{2}$}
\def\Z4{$\mathbb{Z}_{4}$}
\def\cblue{\color{blue}}
\def\cred{\color{red}}
\def\cgr{\color{green}}
\def\cmag{\color{magenta}}


\title {
Isotope Effects and the Negative Thermal Expansion Phenomena in Ice and Water
}

\author{B. I. Min$^{1}$}
\email[e-mail: ]{bimin@postech.ac.kr}
\author{J.-S. Kang$^2$}
\affiliation{
$^1$Department of Physics, Pohang University of Science and Technology,
Pohang 37673, Korea \\
$^2$Department of Physics, The Catholic University of Korea,
    Bucheon 14662, Korea
    }
\date{\today}

\begin{abstract}
H$_2$O is a unique substance with exceptional thermal properties
arising from the subtle interplay between its electronic, phononic,
and structural degrees of freedom.
Of particular interest in H$_2$O are the negative thermal expansion (NTE)
phenomena, observed in its solid phase (ice) at low temperature,
and in its liquid phase (water) near the freezing temperature.
Furthermore, ice and water exhibit the abnormal volume isotope eﬀect (VIE),
where volume expansions occur when replacing H with its heavier isotope, deuterium (D).
In order to capture more conceptual and intuitive understanding
of intriguing NTE and VIE phenomena in ice and water,
we have explored isotope effects in their NTE and melting properties
by employing a type of Born-Oppenheimer-approximation approach
and the Lindemann criterion.
Our findings demonstrate that unusual isotope effects in these phenomena
stem from competition between zero-point-energy
phonons, thermal phonons, and the hydrogen bonding in H$_2$O.
All these components originate from nuclear quantum mechanical (QM) processes,
revealing that QM physics plays a crucial role
in the seemingly classical ice/water systems.

\end{abstract}

\maketitle


\section{Introduction}
\begin{figure}[t]
\centering
\includegraphics[width=8.5cm] {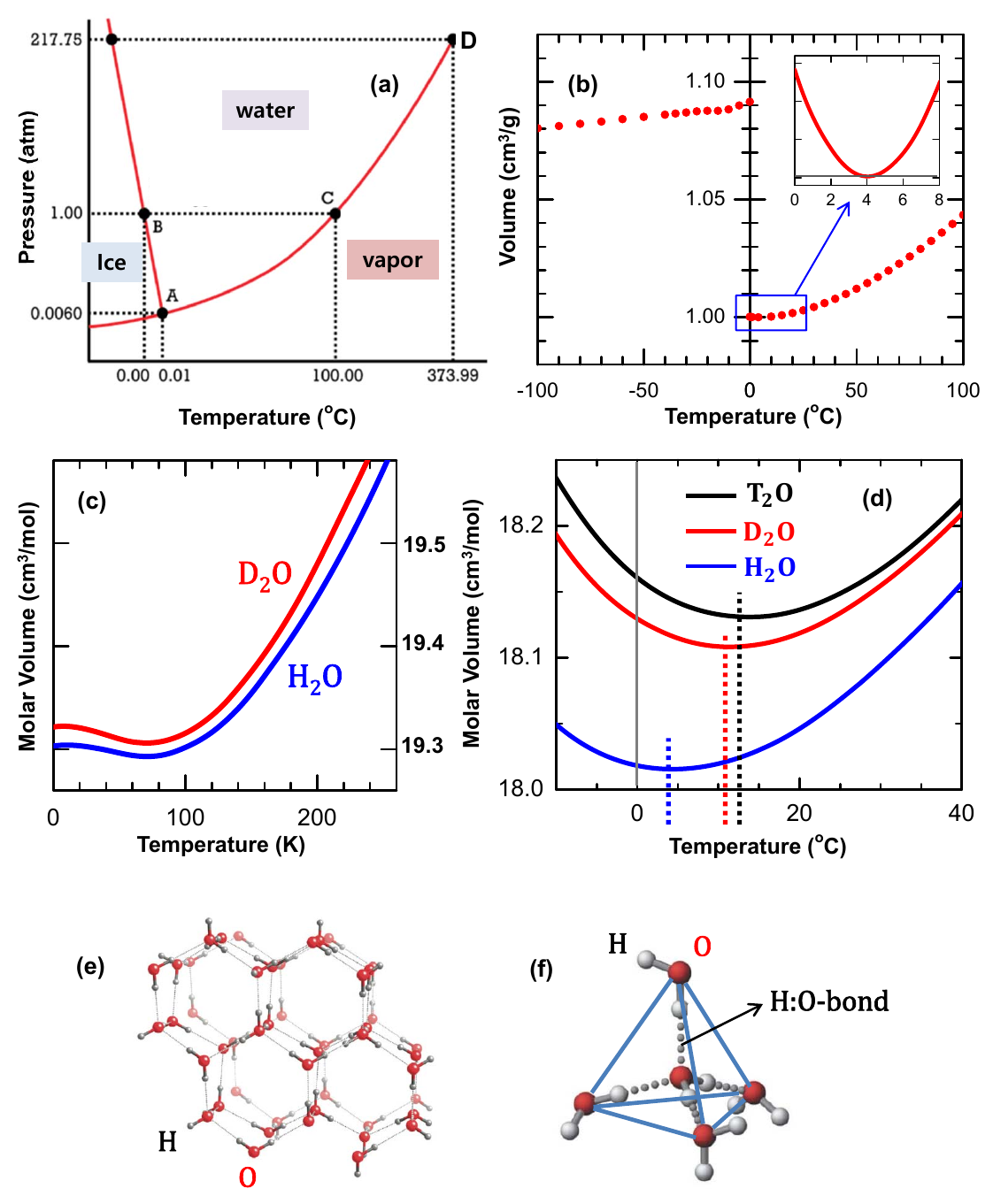}
\caption{
(a) Phase diagram of H$_2$O \cite{PD-h2o}. A, B, C, and D correspond to the triple point,
freezing point, boiling point, and critical point, respectively.
(b) Thermal expansion data near the freezing point \cite{Vol-T}.
Note the NTE (abrupt volume jump)
at the water to ice transition at $T=0~^{\circ}$C (273 K),
and the NTE of water between $0~^{\circ}$C (273 K) and $4~^{\circ}$C (277 K)
 shown in the inset.
(c) The thermal expansion data of ice-Ih phases of H$_2$O and D$_2$O.
The NTEs are observed below $\sim$ 70 K for both.
D$_2$O has the larger volume than H$_2$O,
manifesting the abnormal VIE.
(d) Volume variations of supercooled-water isotopes as a function of $T$
\cite{Water-website,Tmd}.
(e) Open hexagonal crystal structure of ice-Ih (viewed along the $c$ axis).
(f) Hydrogen-bonding (H:O-bonding) in tetrahedral building block of ice-Ih.
Each oxygen forms two-short H-O covalent bonds,
and two-long hydrogen bonds (H:O-bonds) with neighboring H ions.
}
\label{fig1}
\end{figure}

Thermal expansion is a phenomenon of length, area, or volume change of a material
with varying temperature ($T$). Most substances exhibit positive thermal expansions (PTEs),
implying that they expand upon heating and contract upon cooling.
However, H$_2$O in its different solid, liquid, and gas phases
manifests a fascinating exception to this rule
(see the phase diagram in Fig. \ref{fig1}(a)) \cite{PD-h2o}.
Namely, contrary to the general rule, upon cooling, liquid water
expands below $T=4~^{\circ}$C
exhibiting the volume minimum (density maximum) at $T=4~^{\circ}$C,
and then its volume increases abruptly by about 9\% when it freezes
to ice at $T=0~^{\circ}$C ($T$=273 K), as shown in Fig. \ref{fig1}(b).
Hence liquid water has the negative thermal expansion (NTE) coefficient
between $0~^{\circ}$C and $4~^{\circ}$C (inset of Fig. \ref{fig1}(b)),
and ice is less dense than liquid water due to its larger volume \cite{Water-website}.
In fact, the negative slope ($\frac{dP}{dT} <0$)
of phase-equilibrium line between ice and water in Fig. \ref{fig1}(a)
is closely related to the observed NTE at the transition \cite{Liu11,Liu14,Jwkim23,Supp}.
The density maximum at $T=4~^{\circ}$C in H$_2$O is well known to be
essential for ecological systems and various environmental processes \cite{Water-website}.
Furthermore, H$_2$O exhibits a colossal PTE
(volume expands by $\sim$1600 times),
when boiling into its vapor gas phase.

The most common form of ice at the ambient pressure is
the ice-Ih phase \cite{Water-website}.
As shown in Fig.~\ref{fig1}(c), ice-Ih exhibits the normal PTE
between $\sim$70 K and 273 K ($0~^{\circ}$C),
albeit at a much lower thermal-expansion rate than liquid water.
A notable feature in ice-Ih is another NTE phenomenon below $\sim$70 K.
The NTE behavior in ice-Ih at low $T$ is known to arise
from low-energy vibrational modes of ice-Ih phase,
which produce the negative Gr\"{u}neisen parameters
\cite{Liu14,Evans99,Strassle04,Pamuk12,Umemoto15,Gupta18}.

Ice-Ih exhibits an additional abnormal feature in the
volume isotope effect (VIE),
where the volume of the ice changes distinctly depending on the isotope of hydrogen
\cite{Roettger94,Fortes18,Buckingham18}.
In most materials, replacing lighter isotopes with heavier ones leads to a normal VIE,
indicating that the material contracts in volume.
However, in ice-Ih, replacing H with its heavier isotope, deuterium (D),
leads to the volume expansion, as shown in Fig.~\ref{fig1}(c).
This abnormal VIE was reported to arise from the quantum-mechanical (QM) nature
of the hydrogen atoms in the ice lattice \cite{Pamuk12,Umemoto15,Salim16,Cherubini21,Rasti22}.
Interestingly, the abnormal VIE is also observed in water near the so-called
$T_{MD}$ ($T_{MD}$: the temperature corresponding to the maximum density),
as shown in Fig.~\ref{fig1}(d).

The anomalous thermal expansion properties of ice and water
highlight the complex physics and chemistry of the H$_2$O system.
Through decades of research, their mechanisms are understood to some extent
based on numerical computations of the Gr\"{u}neisen parameters of phonon modes
at low $T$ \cite{Pamuk12,Umemoto15,Salim16,Gupta18}
and numerical simulations of molecular-dynamics at high $T$
\cite{Morrone08,Ramirez10,McBride12,Xu20,Eltareb21,Romanelli24}.
Despite extensive studies, however, conceptual understanding
of the underlying physics and chemistry of intriguing NTE and VIE phenomena in ice and water
is still lacking. So we revisit these issues to capture the
comprehensive and intuitive picture from a more fundamental point of view.

In this paper, to identify the physics behind the NTE phenomena in ice and water,
we have examined the isotope effects in the NTE properties,
such as the volume-minimum temperatures ($T_{vm}$'s) and the melting
temperatures ($T_{m}$'s) as well as the VIEs.
Specifically, we have addressed the issues on
(i) the origin of different $T_{vm}$'s for different isotopes
as well as  the abnormal VIE observed in ice-Ih at low $T$,
and (ii) the origin of different $T_{m}$'s and $T_{MD}$'s for different isotopes
as well as the abnormal VIE observed in ice-Ih and water near the freezing $T$.
In the following, the approach we have adopted for the present analysis,
including a type of Born-Oppenheimer approximation
and the Lindemann criterion, is briefly described (Section II),
and then results and discussions of our findings are presented (Section III),
which is followed by concluding remarks (Section IV).

\section{Methodology}

Based on the analysis of existing experimental data for ice and water,
we have attempted to provide conceptual interpretation
of anomalous NTE/VIE and melting properties observed in ice and water.
For this purpose, we have newly employed a type of Born-Oppenheimer
approximation \cite{Marechal68,Marechal07}
to explain the former and the Lindemann criterion \cite{Lindemann1910}
to explain the latter.

A type of Born-Oppenheimer approximation we have employed
is described in the supplement section (ii) \cite{Supp}.
For a system like H$_2$O having the fast-moving high-energy ($\omega$)
and the slow-moving low-energy ($\Omega$) phonon modes, which are far-separated in energy,
it is possible to separate adiabatically the high-energy and low-energy
phonon modes \cite{Marechal68,Marechal07}.
This procedure is just like a conventional Born-Oppenheimer approximation
approach applied for a system with the fast-moving electron
and slow-moving ion motions \cite{AM}.
Then the zero-point energy ($\frac{1}{2}\hbar\omega$) of high-energy phonon mode
plays a role of additional potential energy for the low-energy phonon mode.
We have examined the role of this zero-point energy in the anomalous NTE/VIE properties
of ice and water.

The application of Lindemann criterion to the melting property of ice
is rather rare \cite{With23}.
Here, to explain the unusual isotope effects observed in the melting of ice,
we have adopted a simple Lindemann criterion,
which claims that a crystal melts when the average amplitude of thermal vibrations
(${\langle u^2 \rangle}^{1/2}$) is larger than some fraction of inter-atomic distance.
The derivation of ${\langle u^2 \rangle}$ is provided in
the supplement section (vi) \cite{Supp}.

\section{Results and Discussions}
\subsection{H\lowercase{ydrogen} (H:O)-\lowercase{bonds in ice and water}}

The anomalous thermal expansion behaviors in ice-Ih and water are known to arise
from their characteristic hydrogen-bonding (H:O-bonding) networks \cite{Water-website}.
The H:O-bonding arises from an intermolecular attractive interaction,
in which a H atom that is covalently bonded to a highly electronegative O atom is
attracted to a lone pair of electrons of an O atom in a neighboring molecule.
These intermolecular H:O-bonds, albeit much weaker than
intramolecular H-O covalent bonds, play a crucial role
in determining the arrangement and dynamics of ice  and water.

Ice exists in nearly twenty forms of solid phases,
each with a unique arrangement of H$_2$O molecules
held together by H:O-bonds \cite{Water-website}.
Among those, Ice-Ih crystallizes in an open hexagonal lattice,
in which oxygen ions form an ordered Wurtzite structure
having a low packing ratio, as shown in Fig.~\ref{fig1}(e) \cite{Isaacs99}.
The H:O-bonding network in ice-Ih is composed of tetrahedral building blocks,
where each H$_2$O molecule forms four H:O-bonds with its neighboring molecules,
as depicted in Fig.~\ref{fig1}(f). In ice-Ih, light H ions are not ordered
due to their large zero-point motions.

The NTE observed at low $T$ ($< 70$ K) in ice-Ih is related to its open,
tetrahedral-network structure formed by H:O-bonds.
It has been found that low-energy transverse phonon modes in ice-Ih,
which are associated with bending motions of three hydrogen-bonded molecules,
bring about the negative Gr\"{u}neisen parameter and consequently
the NTE phenomenon \cite{Liu14,Evans99,Strassle04,Pamuk12,Umemoto15,Salim16,Gupta18}.
In fact, such NTE phenomena are manifested in other
tetrahedrally-arranged solids having a low packing ratio, such as Si
and Ge \cite{Liu14,Evans99,Soma77}.
However, they do not exhibit anomalous VIEs but just normal VIEs,
distinct from ice-Ih \cite{Noya97,Herrero99}.

\begin{figure*}[t]
\centering
\includegraphics[width=16cm] {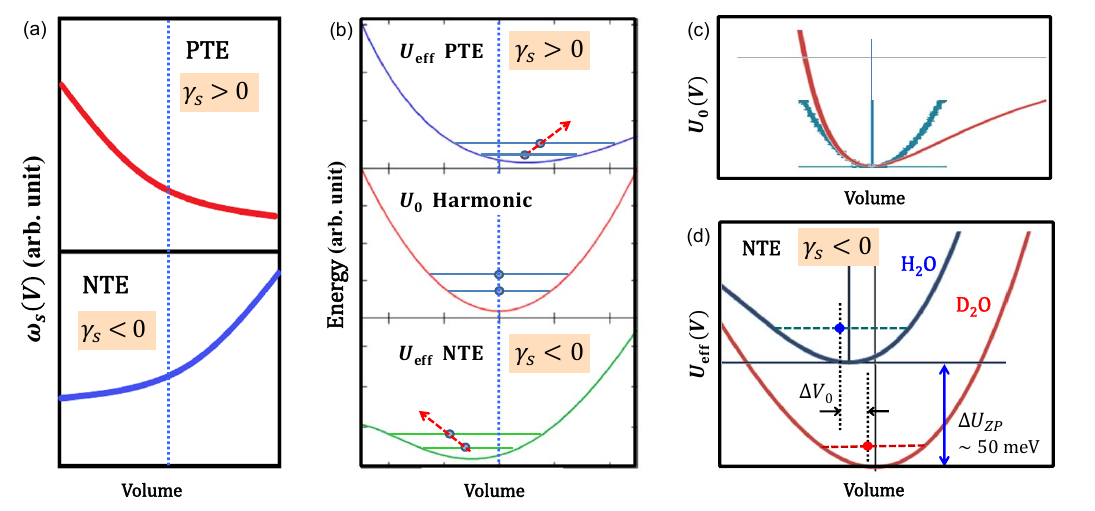}
\caption{(a) Typical volume-dependent behaviors of
high-energy phonon modes $\omega_s(V)$.
The upper one exhibiting the pressure-induced hardening
produces the positive Gr\"{u}neisen parameter and the PTE,
while the lower one exhibiting the pressure-induced softening
produces the negative Gr\"{u}neisen parameter and the NTE.
(b) Effective intermolecular potential energies $U_{eff}$'s
($U_{eff}=U_{0}+U_{ZP}$) for the PTE-anharmonic (top)
and the NTE-anharmonic (bottom) systems, respectively.
$U_{0}$ is assumed to be harmonic (middle).
Red dotted lines in the top and bottom illustrate the volume expansion
and contraction, respectively, with increasing $T$.
(c) The intermolecular potential energy, $U_{0}(V)$ (red curve),
which is intrinsically PTE-anharmonic,
can be approximated by the harmonic potential (blue curve) at low $T$.
(d) For the NTE case with negative $\gamma_s$, the heavier-mass system of D$_2$O ice
has the larger volume (red dot)
than the lighter-mass system of H$_2$O ice (blue dot) ($\Delta V_0 ~(\equiv V_0^{D}-V_0^{H}) > 0$).
Due to the mass-dependent QM zero-point energy $U_{ZP}$,
D$_2$O ice would have larger binding energy than H$_2$O ice by $\Delta U_{ZP}$
($\Delta U_{ZP}$ value taken from Refs. \cite{Rasti22,Pamuk15}).
}
\label{fig2}
\end{figure*}


\subsection{Isotope Effects in the NTE of ice-Ih at low $T$:
Born-Oppenheimer approximation}

In the Born-Oppenheimer approximation,
the adiabatic separation of the fast-moving electron wave-function and the slow-moving ion
wave-function is possible, and then
the obtained volume-dependent eigenvalue of electron ($\epsilon(V)$)
acts as an additional interionic potential energy for the ion motions \cite{AM}.
In a similar way, for a system with the high-energy ($\omega$)
and the low-energy ($\Omega$) phonon modes, which are far-separated in energy,
one can separate adiabatically the high-energy (described by $q$, $p$ variables with mass $m$)
and low-energy (described by $Q$, $P$ variables with mass $M$) phonon modes \cite{Marechal68,Marechal07}.

Then, as described in the supplement \cite{Supp},
the zero-point phonon energy of the fast-moving $q$ mode
obtained in the quasi-harmonic approximation plays a role of the additional potential
energy for the low-energy phonon modes:
\begin{equation}
\left[ \frac{P^2}{2M} + U_0(Q) +
\frac{1}{2}\hbar\omega(Q)\right] \chi_0^N(Q) = E_0^N \chi_0^N(Q),
\label{Qmode}
\end{equation}
where $\chi_0^N(Q)$ is the wave function of low-energy phonon (slow $Q$) mode,
and $U_0(Q)$ is the interionic potential energy.
Here $k$-dependence of $\omega$ is neglected by using the Einstein-phonon approximation.

In the case of hydrogen-bonded systems,
the variable $Q$ corresponds to the intermolecular O$-$O bond length, and so
it can be considered to be a volume ($V$) variable \cite{Marechal68,Marechal07}.
Accordingly, the effective intermolecular potential energy
for the low-energy phonon modes $U_{eff}(V)$ is given by
\begin{equation}
U_{eff}(V) = U_0(V)+U_{ZP}(V),
\end{equation}
where $U_0(V)$ is the intermolecular potential energy
for the static lattice, and $U_{ZP}(V) (\equiv \frac{1}{2} \hbar\omega(V))$
is the QM zero-point energy of the high-energy phonon mode.
Consequently, the low-energy vibrational motions become
significantly affected by $U_{ZP}(V)$,
depending on the volume-dependence of $\omega(V)$,
as provided in Fig.~\ref{fig2} below.

Now consider a system having the high-energy phonon-mode $\omega_s(V)$,
which decreases or increases with volume,
as displayed in Fig.~\ref{fig2}(a).
In fact, $U_0(V)$ for the static lattice is intrinsically anharmonic in nature.
But, at low $T$, when the vibrational amplitudes are small,
$U_0(V)$ can be assumed to be harmonic,  {\it i.e.},
$U_0(V) \approx \frac{1}{2} K(V-V_0)^2$,
as shown in the middle of Fig.~\ref{fig2}(b) and Fig.~\ref{fig2}(c).
Then, through the contribution of $U_{ZP}(V)$,
$U_{eff}(V)$ would have either the anharmonic potential
of shallower-flattened-right (top)
or that of shallower-flattened-left (bottom)
with respect to the harmonic potential in the middle.
According to the definition of Gr\"{u}neisen parameter
$\gamma~(\equiv -\partial \ln \omega/\partial \ln V)$ \cite{Supp},
the former describes the PTE system with positive $\gamma_s$,
while the latter describes the NTE system
with negative $\gamma_s$.
The red arrows in the top and bottom of Fig.~\ref{fig2}(b) illustrate
the positive and negative volume expansions, respectively,
with increasing $T$.

In the case of ice-Ih,
the high-energy phonon mode corresponds to
the intramolecular O-H stretching mode with energy of $\sim$0.4 eV,
while the low-energy phonon modes correspond to
the intermolecular vibration modes with energy of $\leq$ 0.05 eV,
which are considered to be responsible for the NTE at low $T$
\cite{Umemoto15,Yoshimura11,Sun13,Zhang15}.
Hence the application of the Born-Oppenheimer approximation
would be sufficiently valid in the present case \cite{Libron}.

Note that the high-energy stretching mode in ice-Ih, $\omega_s$, was
observed to possess a pressure-induced phonon softening \cite{Yoshimura11}.
This feature is understood in view of the fact that
the O-H bond lengths tend to increase with increasing $P$.
In fact, its Gr\"{u}neisen parameter was calculated to be
negative ($\gamma_s \approx -0.2)$ \cite{Pamuk12,Pamuk15}.
Then, as described in the bottom of Fig.~\ref{fig2}(b),
the zero-point energy of $\omega_s(V)$ yields an NTE-anharmonic $U_{eff}(V)$.
$U_0(V)$, being an electrostatic interaction,
would be independent of isotopic mass, and so
$U_0(V)$ is same for both H$_2$O and D$_2$O.
Hence, considering the mass-dependent $U_{ZP}$
($\equiv \frac{1}{2}\hbar\omega_s \sim \frac{1}{\sqrt{m_{iso}}}$), D$_2$O ice would have
a higher binding energy than H$_2$O ice by $\Delta U_{ZP}~(\approx50$ meV)
\cite{Rasti22,Pamuk15}, as displayed schematically in Fig. \ref{fig2}(d).

In addition, because of the mass-dependent $U_{ZP}(V)$,
$U_{eff}(V)$ of D$_2$O becomes less anharmonic than
that of H$_2$O, that is, $U_{eff}(V)$ of D$_2$O is closer to
the harmonic potential than that of H$_2$O \cite{slope}.
For simplicity, let us assume that there is just a single low-energy phonon mode,
corresponding to an intermolecular vibration in ice-Ih,
in the presence of $U_{eff}(V)$.
Then, as shown in Fig. \ref{fig2}(d),
H$_2$O having a stronger contribution from the anharmonic NTE potential
would have smaller equilibrium volume (blue dot) than D$_2$O (red dot)
having a weaker contribution from the anharmonic NTE potential.
Thus the asymmetric nature of $U_{eff}(V)$
of an NTE anharmonic type explains why ice-Ih exhibits the abnormal VIE,
namely, D$_2$O ice has the larger equilibrium volume than H$_2$O ice
($V_0^{D} > V_0^{H}$ at $T=0$).

The volume at the minimum of $U_{eff}(V)$, $V_m$, can be obtained quantitatively
from the volume derivative of $U_{eff}(V)$.
Considering the expansion of $\frac{1}{2}\hbar\omega_s(V)$ up to the first order,
$U_{eff}(V) = \frac{K}{2}(V-V_0)^2 + \frac{1}{2}\hbar\omega_s(V)
   \approx  \frac{K}{2}(V-V_0)^2 +  \frac{1}{2}\hbar\omega_s(V_0)
    + \frac{\hbar}{2} (V-V_0) \frac{\partial \omega_s}{\partial V}|_{V_0}$.
This yields
$V_m=V_0- \frac{\hbar}{2K} \frac{\partial  \omega_s}{\partial V}|_{V_0}$
from $\frac{\partial U_{eff}(V)}{\partial V}|_{V_m}=0$.
Therefore, the isotopic volume difference between the D$_2$O and H$_2$O ice,
$\Delta V_m~(\equiv V_m^{D}- V_m^{H})$, is approximately given by
\begin{equation}
\Delta V_m \approx \frac{\hbar}{2K} \left(\frac{\partial \omega_s^{H}}
    {\partial V}-\frac{\partial \omega_s^{D})}{\partial V}\right)
    \approx \frac{\hbar}{2K} \frac{\partial \omega_s^{H}}{\partial V}
        \left( 1-\sqrt{m_{H}/m_{D}}\right),
\end{equation}
which is positive because $\frac{\partial \omega_s^{H}}{\partial V}>0$,
implying the abnormal VIE.

Indeed, the observed NTE for both H$_2$O and D$_2$O below $\sim$70 K in Fig. \ref{fig1}(c) is
explained well by the NTE-anharmonic $U_{eff}$ shown at the bottom of Fig. \ref{fig2}(b).
But, as $T$ increases above $\sim$70 K, the PTE is observed for both H$_2$O and D$_2$O,
as if $U_{eff}(V)$ changes from the NTE-anharmonic to the PTE-anharmonic type.
This feature can be understood based on the change of $U_0(V)$ from the
harmonic to the anharmonic form with increasing $T$, as shown in Fig. \ref{fig2}(c).
Remember that, in Fig. \ref{fig2}(d), the anharmonic $U_0(V)$ was assumed
to be harmonic at low $T$ \cite{ice-III}.
But, with increasing $T$, the intrinsically PTE-anharmonic nature of $U_0(V)$ starts
to be more and more effective,
and so $U_{eff}(V)$ of ice-Ih eventually becomes PTE-anharmonic,
leading to volume expansions for both H$_2$O and D$_2$O.
Namely, with increasing $T$, the phonon modes with positive Gr\"{u}neisen parameters
become populated, and thereby reduce the NTE.
Interestingly, even at finite $T$ above $\sim$70 K,
the abnormal VIE will be maintained because
$U_{eff}(V)$ of H$_2$O is still shallower-fattened-left than that of D$_2$O
due to a mass-dependent $U_{ZP}$ contribution,
and thereby the volume of H$_2$O is still smaller than that of D$_2$O.

As shown in Fig.~\ref{fig1}(c), there is also an isotope effect
on the volume-minimum temperatures ($T_{vm}$), albeit not so prominent.
In fact, $T_{vm}$ in H$_2$O ice is only slightly higher ($\sim$1 K)
than that in D$_2$O ice \cite{Roettger94, Buckingham18}.
$T_{vm}$ would be determined by the crossing point at which
$U_{eff}(V)$ changes from an NTE-anharmonic type to a PTE-anharmonic type.
As described above, the NTE-anharmonic feature will persist
up to higher $T$ in H$_2$O than in D$_2$O, and so
$T_{vm}$ in H$_2$O is higher than $T_{vm}$ in D$_2$O
($\sim$72 K for H$_2$O vs. $\sim$71 K for D$_2$O) \cite{Roettger94}.

\subsection{Isotope Effects in the NTE of water near the freezing (melting) $T$}

As $T$ increases, the thermal vibrational energy of the H$_2$O molecules in ice-Ih
increases, which weakens the H:O-bonds and causes the molecules to fluctuate
around their equilibrium positions.
Then H:O-bonds become weaker and weaker so as to be broken,
and eventually the volume collapse occurs at the melting temperature
$T_m=0~^{\circ}$C.
This feature of ice-Ih is different from those of most solids,
for which the extra molecular motion in the liquid phase requires more
space and therefore melting is accompanied by volume expansion.

As shown in Fig. \ref{fig1}(b),
H$_2$O water has the minimum volume at $T_{MD}=4~^{\circ}$C.
This volume minimum is brought about by the balance
between two competing effects with increasing $T$:
the volume contraction driven by melting vs. the normal volume expansion.
As mentioned above, the crystalline H:O-bonding network is to be
broken at $T_m=0~^{\circ}$C.
In the liquid phase, however, substantial amounts of H:O-bonds are still maintained
in cluster forms of hexagonal ice-Ih phase keeping large empty space \cite{Nemethy62}.
Upon heating, the number of these remaining H:O-bonds decreases slowly,
resulting in the volume contraction.
Meanwhile, due to the increased kinetic vibrational energy
of unbonded H$_2$O molecules, water expands upon heating like most substances.
The crossing of these two effects is supposed to occur at $T_{MD}$.

As introduced in Fig. \ref{fig1}(d),
the isotope effects are also manifested in water near $T_{MD}$.
As provided in Table I,
D$_2$O water has significantly higher $T_{MD}$ ($\sim11.2~^{\circ}$C)
and the higher freezing (melting) temperature $T_{m}$ ($\sim3.7~^{\circ}$C)
than H$_2$O.
Heavier T$_2$O (T: tritium) water has even higher $T_{MD}$
and $T_{m}$ \cite{Tmd,McBride12}.
Thus D$_2$O and T$_2$O waters exhibit wider temperature ranges of the NTE phenomena
($7.5$ and $8.9~^{\circ}$C)
than H$_2$O water ($4~^{\circ}$C),

\begin{table}[b]
\caption{
Isotope effects on the freezing (melting) temperature ($T_{m}$) and the
maximum-density temperature ($T_{MD}$) \cite{McBride12}.
Here H, D, and T represent hydrogen, deuterium, and tritium, respectively.
}
\begin{ruledtabular}
\centering
\begin{tabular}{l c c c c c c}
            & $T_{m}$ [K] &  [$^{\circ}$C] &  $T_{MD}$ [K] &  [$^{\circ}$C]  & $(T_{MD}-T_{m})$ [K]  \\
\hline
H$_2$O      & 273.15  & 0    & 277.13   & 3.98  & 3.98   \\
D$_2$O      & 276.83  & 3.68 & 284.34   & 11.19 & 7.51   \\
T$_2$O      & 277.64  & 4.49 & 286.55   & 13.40 & 8.91   \\
\end{tabular}
\label{iso}
\end{ruledtabular}
\end{table}

Such differences are expected to occur due to
the competition between the isotopic mass-dependent H:O-bonding ($E_{HB}$)
and the kinetic molecular-vibration ($E_k$) energies
of H$_2$O and D$_2$O.
$E_k$ increases upon heating, and
melting occurs when $E_k$ overcomes
the intermolecular energy holding the crystal structure together
that corresponds to the cohesive energy ($E_{coh}$) \cite{Deeney08}.
When H atoms are replaced with D atoms,
the molecules would have stronger intermolecular D:O-bondings
due to the reduced zero-point QM fluctuations of heavier D ions.
On the other hand,
larger intramolecular quantum delocalization of H ions in H$_2$O brings about
the enhanced dipole moment of the water molecule, producing
a stronger H:O-bonding network in H$_2$O than in D$_2$O.
Hence the energy difference in H:O and D:O-bonding energies will be much reduced
due to two opposite effects of intermolecular and intramolecular
nuclear QM fluctuations \cite{Habershon09}.
Nevertheless, the D:O-bonding network is expected to be still
more robust than the H:O-bonding network \cite{McBride12,Ceriotti16},
and so $E_{coh}$ of D$_2$O is larger than that
of H$_2$O by their difference in $E_{HB}$ ($\Delta E_{HB}$).
$E_{coh}$ of H$_2$O ice was reported
to be $\sim$600 meV experimentally \cite{Whalley57,Whalley84},
and $E_{coh}$ of D$_2$O was estimated to be larger than $E_{coh}$ of H$_2$O
by $\Delta E_{HB}$ of $\sim$15 meV \cite{Pamuk15}.
Meanwhile, $E_k$ including the zero-point motions
in H$_2$O ice was reported to be as much as $\sim$350 meV at low $T$,
whereas $E_k$ in D$_2$O is supposed to be smaller than $E_k$ in H$_2$O
by $\Delta E_k \sim$90 meV due to its heavier mass \cite{Romanelli24}.
Both $E_k$'s increase slowly with increasing $T$.

\begin{figure}[t]
\includegraphics[width=8.5cm] {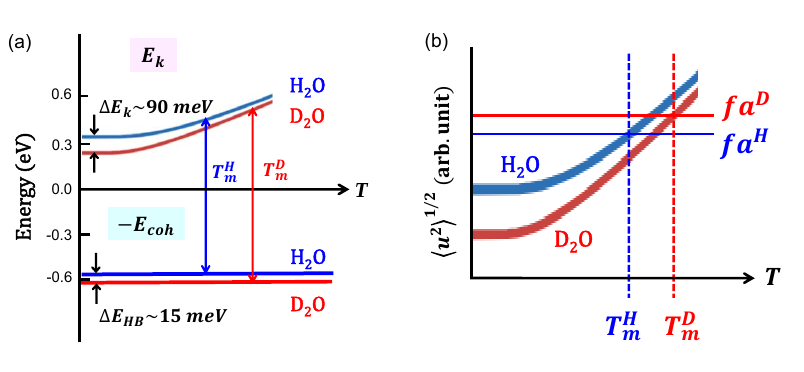}
\caption{
(a) Schematic plot of $T$-dependent variations of the
kinetic molecular vibration energies ($E_k$'s)
and the ($-1\times$) cohesive energies ($-E_{coh}$'s) of H$_2$O (in blue) and D$_2$O (in red) ices.
The variation of $E_k$ is somewhat exaggerated.
Melting occurs when $E_k$ overcomes $E_{coh}$.
It is shown that $T_{m}$ of D$_2$O is higher than that of H$_2$O.
$\Delta E_k$ and $\Delta E_{HB}$ values are taken from Ref. \cite{Romanelli24}
and Ref. \cite{Pamuk15}, respectively.
(b) Thermal-average amplitudes of ionic displacements ${\langle u^2 \rangle}^{1/2}$
of H$_2$O (blue) and D$_2$O (red) ices.
Melting occurs when ${\langle u^2 \rangle}^{1/2} \geq fa$
($a$: intermolecular distance, $f$: fractional number ($\leq 0.2)$).
}
\label{fig3}
\end{figure}

In Fig. \ref{fig3}(a) are plotted $E_k$'s and $-E_{coh}$'s
of H$_2$O and D$_2$O ices as a function of $T$.
It is shown that the larger $E_k$ in H$_2$O ice
will break the weaker H:O-bonds in H$_2$O ice at a lower $T$ than in D$_2$O ice,
resulting in the lower $T_{m}$ for H$_2$O ice than for D$_2$O ice
($T^H_m < T^D_m$).
Also, as a consequence, the lower $T_{MD}$ is naturally expected
for H$_2$O water than for D$_2$O water.

This feature can also be analyzed by using the Lindemann criterion \cite{Lindemann1910}.
According to this criterion, a crystal melts
when the average amplitude of thermal vibrations of atoms
(${\langle u^2 \rangle}^{1/2}$)
is larger than some fraction of interatomic distances ($fa$), {\it i.e.},
\begin{equation}
{\langle u^2 \rangle}^{1/2} \geq fa,
\end{equation}
where $u$ is the atomic displacement, $f$ is some fractional number
called the Lindemann parameter ($f \leq 0.2$ in general),
and $a$ is the interatomic distance.
As described in the supplement \cite{Supp}, $\langle u_i^2 \rangle$ for $i$-th atom is given by
$\sum_{k,\lambda} \frac{\hbar}{2M_i\omega_{k\lambda}} \coth{\frac{\beta \hbar\omega_{k\lambda}}{2}}$
($\beta=\frac{1}{k_B T}$),
indicating that all the phonon modes contribute to $\langle u_i^2 \rangle$.
Note that $\langle u_i^2 \rangle$ has isotopic-mass dependence through $M_i$ and $\omega_{q\lambda}$,
and, at $T=0$ K, $\langle u_i^2 \rangle = \frac{\hbar}{2M_i\omega_{q\lambda}}$.
To examine the isotope effect on $T_{m}$,
let us consider just one specific phonon mode ($q\lambda$) for simplicity,
$\langle u_i^2 \rangle=\frac{\hbar}{2M_i\omega_{q\lambda}}
\coth{\frac{\beta \hbar\omega_{q\lambda}}{2}}$.
Accordingly, $T_{m}$ can be obtained from
\begin{equation}
{\langle u_i^2 \rangle}^{1/2} = \left[\frac{\hbar}{2M_i\omega_{q\lambda}}
\coth{\frac{\beta_m \hbar\omega_{q\lambda}}{2}}\right]^{1/2}=fa_i,
\label{tm}
\end{equation}
as schematically plotted in Fig. \ref{fig3}(b).
Due to a heavier mass of D$_2$O than that of H$_2$O,
${\langle u_i^2 \rangle}^{1/2}$ of D$_2$O  is to be smaller than that of H$_2$O.
On the other hand, since the volume of D$_2$O is larger than that of H$_2$O,
$fa_i$ of D$_2$O is to be larger than that of H$_2$O. $T_{m}$ is obtained from
the crossing of these two graphs, and so $T^D_{m}$ of D$_2$O is to be higher
than $T^H_{m}$ of H$_2$O.

\begin{figure}[t]
\includegraphics[width=8.5cm] {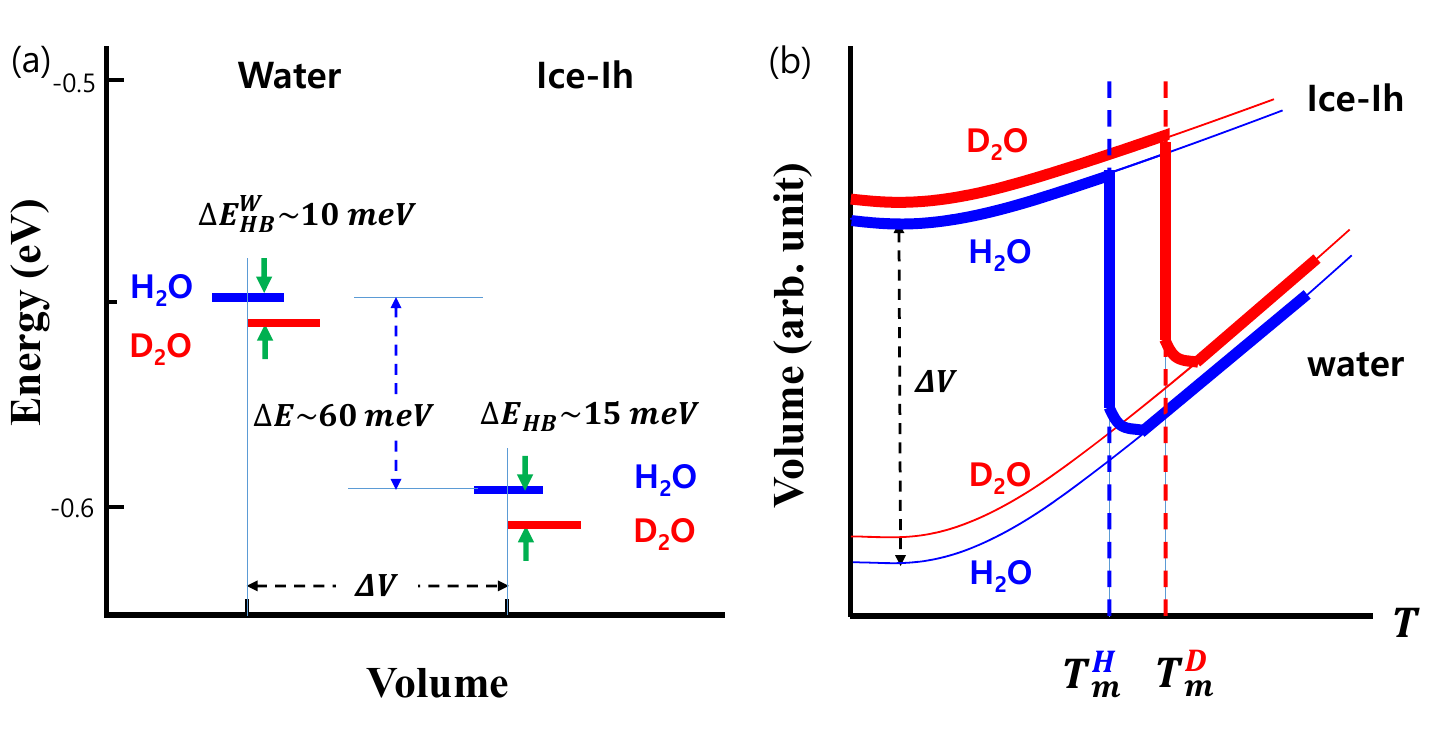}
\caption{
(a)
Energy configurations of H$_2$O and D$_2$O waters and ice-Ih's.
The nuclear QM effects are still retained in the liquid phase near $T_m$, and so
D$_2$O water has the larger H:O-bonding energy $E^{W}_{HB}$ and the larger volume
than H$_2$O water, as in ice-Ih phase.
The value of $\Delta E$ is taken from Refs. \cite{Ramirez10,Nemethy62},
and that of $E^{W}_{HB}$ is taken from Ref. \cite{Hakem07}.
(b) Thermal expansion behaviors of H$_2$O and D$_2$O ice-Ih's and waters
expected from the energy configurations of (a).
}
\label{fig4}
\end{figure}


Let us try to estimate $T_{m}$ for a specific phonon mode $\omega_{q\lambda}$.
The intermolecular H:O-bond stretching vibration mode of H$_2$O ice is
expected to be the most relevant mode
to the melting of ice-Ih \cite{Umemoto15,Gupta18,Rasti22}.
So let us select $\omega^H_{q\lambda}=215$ cm$^{-1}$ ($\approx 27$ meV) for H$_2$O,
which is close to a center of phonon density of states
of H:O-bond stretching vibration modes.
%
For D$_2$O, let us select $\omega^D_{q\lambda}=204$ cm$^{-1}$ ($\approx 25$ meV),
which is determined
by counting the molecular-mass difference between D$_2$O and H$_2$O
($\omega^D_{q\lambda}=\omega^H_{q\lambda}\sqrt{M_{H_2O}/M_{D_2O}}$).
Then, in the case of H$_2$O, $T^H_m$ is obtained to be 274.6 K,
for $f=0.2$, $m_{H_2O}=18$ g/mole, and $a^H=2.76~ \AA$ \cite{Giubertoni23} in Eq. (\ref{tm}).
Here $a^H$ corresponds to the intermolecular O$-$O distance.
Similarly, in the case of D$_2$O, $T^D_m$ is obtained to be 280.4 K,
for $f=0.2$, $M_{D_2O}=20$ g/mole, and $a^D=2.80~ \AA$ \cite{Giubertoni23} in Eq. (\ref{tm}).
Therefore, the higher $T^D_{m}$ of D$_2$O than $T^H_{m}$ of H$_2$O is
elucidated well, and the intermolecular bond-stretching vibration modes
are suggested to play an important role in the melting of ice-Ih.

%

As displayed in Fig. \ref{fig1}(d),
water also exhibits the abnormal VIE, namely, larger volume for D$_2$O water
than for H$_2$O water.
As mentioned above, near $T_{MD}$, substantial amounts of H:O-bonds in water
remain in the ice-like clusters, which are embedded in and in equilibrium with
the unbonded water molecules \cite{Nemethy62}.
This implies that the nuclear QM effects, the hallmark in ice-Ih phase, are retained
even in liquid phases above the melting temperature.
Then one can understand the abnormal VIE in water,
by employing again the $U_{ZP}$-induced NTE-anharmonic intermolecular potential energy,
similarly as discussed in Fig. \ref{fig2}(d) for ice-Ih.
That is, $U_{eff}$'s of water are quite similar in shape to those of ice-Ih,
so that H$_2$O having a larger contribution from the anharmonic NTE potential
would have a smaller equilibrium volume
than D$_2$O having a smaller contribution
from the anharmonic NTE potential.

Further, due to existing nuclear QM effects in liquid phases,
D$_2$O and H$_2$O waters would have different H:O bonding energies, $E^W_{HB}$'s.
Figure \ref{fig4} summarizes the energy configurations and
thermal expansion behaviors of D$_2$O and H$_2$O waters and ice-Ih's.
Figure \ref{fig4}(a), constructed based on the data in the literature, provides the following
informations:
(i) D$_2$O has larger volume than H$_2$O for both ice-Ih and water,
(ii) waters are located at a higher energy (by $\Delta E \sim$60 meV) \cite{Ramirez10,Nemethy62} and at a smaller volume than those of ice-Ih,
(iii) D$_2$O and H$_2$O have different H:O bonding energies,
$\Delta E_{HB}\sim 15$ meV for ice-Ih \cite{Pamuk15}
and $\Delta E^W_{HB}\sim 10$ meV for water \cite{Hakem07}.

In Fig. \ref{fig4}(b), $T$-dependent thermal expansion behaviors
expected from the energy configurations given in Fig. \ref{fig4}(a) are provided.
It is shown that
D$_2$O possesses the larger volume for both ice-Ih and water
phases and the higher $T_m$ than for H$_2$O.


\section{Concluding Remarks}

Prominent isotope effects observed in ice and water indicate that
the nuclear QM nature of light H and D ions plays an important role in their thermal physics.
Indeed, by employing a type of Born-Oppenheimer approximation
for ice-Ih, which has very high-energy phonon modes
arising from intramolecular H-O and D-O bond stretching vibrations,
we are able to capture a conceptual and intuitive picture
of their intriguing NTE and VIE phenomena in the same framework.
We demonstrate
that the abnormal VIE phenomena in ice and water are attributed
to the NTE-anharmonic $U_{eff}$ driven by the QM zero-point energy of the high-energy
intramolecular stretching phonon modes with negative $\gamma$.
Furthermore, it is found that isotope effects on $T_{vm}$ and $T_{m}$ of ice,
as well as $T_{MD}$ of water,
originate from the enhanced nuclear QM fluctuations in H$_2$O compared to D$_2$O.
{\cred
}

These findings reveal that the interplay between
zero-point energy phonons, thermal phonons,
and the H:O-bondings, all of which are of QM origin, governs
the unique NTE, VIE, and melting behaviors of ice and water.
A comprehensive and conceptual understanding of these mechanisms
would provide valuable insights into not only the NTE
but also other hitherto-mysterious thermal properties observed in ice and water,
including specific-heat, viscosity, compressibility, and sound velocity. \\ \\


{\bf Acknowledgments:}

Helpful discussions with Prof. Kwang-Sei Lee and Prof. Byeong June Min are greatly appreciated.
JSK acknowledges the support by the National Research
Foundation (NRF) of Korea (Grant No. RS-2023-00275779). \\  \\



\clearpage


{\bf {{\it Supplementary Material}:
``Isotope Effects and the Negative Thermal Expansion Phenomena in Ice and Water"}} \\ \\


%



\renewcommand{\thefigure}{S\arabic{figure}}
\setcounter{figure}{0}
\renewcommand{\thetable}{S\arabic{table}}
\setcounter{table}{0}
\renewcommand{\thesection}{S\arabic{section}}
\setcounter{section}{0}
\renewcommand{\theequation}{S\arabic{equation}}
\setcounter{equation}{0}

\renewcommand*{\citenumfont}[1]{S#1}
\renewcommand*{\bibnumfmt}[1]{[S#1]}




{\bf (i) Negative thermal expansion (NTE) when $\frac{dP}{dT} <0$}

The negative slope ($\frac{dP}{dT} <0$)
of phase-equilibrium line at the ice-water transition in Fig. 1(a)
of the main text
suggests the NTE behavior across the ice-water transition upon heating.
It is because the thermal-expansion coefficient $\a_V$ is given by
\begin{equation}
\a_V=\frac{1}{V} \left(\frac{\partial V}{\partial T}\right)_P
    =\frac{1}{B} \left(\frac{\partial P}{\partial T}\right)_V,
\label{alphav1}
\end{equation}
where $B$ is the bulk-modulus,
$B=-V \left(\frac{\partial P}{\partial V}\right)_T$,
which is always positive \cite{SAM}.
Here we used the thermodynamic relation,
$
\left(\frac{\partial V}{\partial T}\right)_P=-\frac{(\partial P/\partial T)_V}
{(\partial P/\partial V)_T}.
$
Accordingly, $\a_V$ can be negative, provided
$\frac{dP}{dT}~<~0$ in the pressure-temperature ($PT$) phase diagram. \\

{\bf (ii) Adiabatic separation of the high- and low-energy phonon modes}

Let us consider a system having the high-energy and low-energy phonon modes,
which are far-separated in energy.
Then, as done in the Born-Oppenheimer approximation for the fast electron and slow ion motions \cite{SAM},
one can separate adiabatically the high-energy (fast $q$) and low-energy (slow $Q$) phonon modes.

Consider the total Hamiltonian for $q$ (with mass $m$) and $Q$ (with mass $M$) ion motions:
\begin{equation}
H(q,Q) \Psi_n(q,Q) = E_n(q,Q) \Psi_n(q,Q),
\label{totph}
\end{equation}
where
\begin{equation}
H(q,Q)=\frac{p^2}{2m} + \frac{P^2}{2M} + U(q,Q).
\end{equation}
Here $U(q,Q)$ corresponds to the static inter-ionic potential obtained by using
the Born-Oppenheimer approximation.
In the harmonic approximation for $U(q,Q)$,
\begin{eqnarray}
H(q,Q) &=& \frac{p^2}{2m} + \frac{1}{2}m\omega(Q)^2(q-q_0)^2  \nonumber \\
       &+& \frac{P^2}{2M} + \frac{1}{2}M\Omega^2(Q-Q_0)^2.
\end{eqnarray}

Now consider the adiabatic separation of fast ($q$) and slow ($Q$) ionic motions,
\begin{equation}
\Psi_n(q,Q)=\chi_n(Q)\phi_n(q,Q)
\label{Psi}.
\end{equation}
Here $\phi_n(q,Q)$ is the wave function of high-energy phonon (fast $q$) mode, satisfying
\begin{equation}
 \left[ \frac{p^2}{2m} + \frac{1}{2}m\omega(Q)^2(q-q_0)^2  \right] \phi_n(q,Q) =
\left( n+\frac{1}{2} \right)\hbar\omega(Q)\phi_n(q,Q).
\label{phi}
\end{equation}
Here $Q$ is considered just as a given parameter.

Substituting Eq. (\ref{Psi}) and (\ref{phi}) into Eq. (\ref{totph}), and using the
adiabatic approximation \cite{SMarechal68,SMarechal07},
then we have the following equation for the low-energy phonon mode,
\begin{equation}
\left[ \frac{P^2}{2M} + \frac{1}{2}M\Omega^2(Q-Q_0)^2 +
\frac{1}{2}\hbar\omega(Q)\right] \chi_0^N(Q) = E_0^N \chi_0^N(Q).
\label{SQmode}
\end{equation}
In Eq. (\ref{SQmode}), we considered the $n=0$ ground-state eigenvalue of $q$ phonon mode,
and so the effective potential for $Q$ phonons is given by
\begin{equation}
U_{eff}(Q) = \frac{1}{2}M\Omega^2(Q-Q_0)^2 + \frac{1}{2}\hbar\omega(Q)
\label{Ueff}.
\end{equation}
Therefore, in this adiabatic approximation, the zero-point
phonon energy of fast motion of $q$ mode ($U_{ZP}\equiv \frac{1}{2}\hbar\omega(Q)$)
plays a role of additional potential energy for the slow motion of $Q$ modes.
In the case of H$_2$O ice and water, the intramolecular O-H stretching mode corresponds
to the fast $q$ mode, while intermolecular vibrations correspond
to the slow $Q$ mode. \\

{\bf (iii) Thermal expansion coefficient of phonon origin}

In the quasi-harmonic approximation,
pressure exerted by phonons can be obtained from the
volume-dependent free energy of phonons, $F_{ph}(V)$,
by using $P=-(\partial F_{ph}(V)/\partial V)_T$  \cite{SAM,SDjkim95,SPamuk12}:
\begin{equation}
F_{ph}(V)=\sum_k \frac{1}{2} \hbar\omega_k(V) + k_B T \sum_k \ln[1-\exp(-\hbar\omega_k(V)/k_B T)].
\end{equation}
The first term corresponds to the energy of the quantum mechanical zero-point vibrations
of the normal modes.
The second term corresponds to the thermal phonon energy at $T\neq0$.
Then
\begin{equation}
P=-\frac{\partial}{\partial V} \left(\sum_k \frac{1}{2} \hbar\omega_k(V) \right) +
    \sum_k \left(-\frac{\partial}{\partial V}\hbar\omega_k(V) \right) n_k,
\label{pressure}
\end{equation}
where $n_k=\frac{1}{e^{\beta \hbar\omega_k}-1}$ is the boson distribution function \cite{SAM}.

Substituting Eq. (\ref{pressure}) into Eq. (\ref{alphav1}),
the thermal-expansion coefficient $\a_V$ of phonon origin is given by
\begin{equation}
\a_V = \frac{1}{B} \sum_k \left(-\frac{\partial}{\partial V}\hbar\omega_k \right)
    \frac{\partial n_k}{\partial T}.
\end{equation}
Now let us define the $k$-dependent Gr\"{u}neisen parameter $\gamma_k$
and the $k$-dependent specific heat $c_v(k)$
contributed by a phonon mode with the energy $\hbar\omega(k)$
(total specific heat $c_v= \sum_k c_v(k)$), as follows:
\begin{eqnarray}
\gamma_k \equiv -\frac{V}{\omega_k}\frac{\partial \omega_k}{\partial V}, \label{gammak}\\
c_v(k) \equiv \frac{1}{V} \hbar\omega_k \frac{\partial n_k}{\partial T}.
\end{eqnarray}
Let us also define the overall Gr\"{u}neisen parameter $\gamma$
corresponding to the weighted average of $\gamma_k$,
\begin{equation}
\gamma \equiv \sum_k \gamma_k c_v(k)/\sum_k c_v(k).
\label{gamma}
\end{equation}

Then $\a_V$ can be expressed as
\begin{equation}
\a_V=\frac{1}{V} \left(\frac{\partial V}{\partial T}\right)_P
    =\frac{\gamma}{B} c_v.
\label{alphav2}
\end{equation}
Thus, whether a system is the NTE or PTE is determined by the sign of $\gamma$. \\

{\bf (iv) Volume change and the Gr\"{u}neisen parameter}

From Eq. (\ref{alphav2}),
\begin{eqnarray}
\frac{dV}{V}&=&\frac{\gamma}{B} c_v dT,\\
\int_{V_0}^{V} \frac{dV}{V}&=& ln \frac{V}{V_0} = \frac{1}{B}\int {\gamma c_v} dT,
\label{vv01}
\end{eqnarray}
and using Eq. (\ref{gamma}),
\begin{eqnarray}
\int \gamma c_v dT &=& \int \sum_k \gamma_k c_v(k) dT \nonumber \\
        &=& \int \sum_k \gamma_k \frac{1}{V_0} \hbar\omega_k \frac{\partial n_k}{\partial T} dT.
\label{gammacv}
\end{eqnarray}

Note that, for $V=V_0+\Delta V$,
\begin{equation}
ln \frac{V}{V_0} = ln \left(1+ \frac{\Delta V}{V_0}\right) \approx \frac{\Delta V}{V_0}.
\label{vv02}
\end{equation}
Accordingly, from Eqs. (\ref{vv01}), (\ref{gammacv}), and (\ref{vv02}),
\begin{equation}
\frac{\Delta V}{V_0}=\frac{1}{B} \sum_k \gamma_k \frac{1}{V_0} \hbar\omega_k n_k.
\end{equation} \\

{\bf (v) Phonon frequency and the Gr\"{u}neisen parameter}

On the other hand, from Eq. (\ref{gammak}),
\begin{eqnarray}
\gamma_k ln \frac{V}{V_0} &=& -ln \frac{\omega_k(V)}{\omega_k(V_0)}, \\
\omega_k(V) &=& \omega_k(V_0)(\frac{V}{V_0})^{-\gamma_k}.
\label{omegakv}
\end{eqnarray}

Eq. (\ref{omegakv}) indicates that, for the NTE with negative $\gamma_k$,
phonon frequency increases with volume (pressure-induced softening),
while, for the PTE with positive $\gamma_k$,
phonon frequency decreases with volume (pressure-induced hardening)
(see Ref. \cite{SPamuk12}). \\


{\bf (vi) Average amplitude of thermally vibrating atoms: $\langle u^2 \rangle^{1/2}$}

For a single harmonic oscillator in one dimension,
the atomic displacement $u_i$ can be described by the quantized creation and
annihilation operators, $a$ and $a^{\dag}$, as follows:
\begin{equation}
u_i=\sqrt{\frac{\hbar}{2M_i\omega}}(a+a^{\dag}).
\end{equation}
Then
\begin{eqnarray}
\langle u_i^2 \rangle &=&
  \frac{\hbar}{2M_i\omega}  \langle (a+a^{\dag}) (a+a^{\dag}) \rangle  \nonumber \\
&=& \frac{\hbar}{2M_i\omega} \langle a a^{\dag} + a^{\dag}a  \rangle  \nonumber \\
&=& \frac{\hbar}{2M_i\omega}  (\langle 2a^{\dag}a \rangle +1)  \nonumber \\
&=& \frac{\hbar}{M_i\omega}  (n +1/2),
\label{ui2-1D}
\end{eqnarray}
where $n ~ (\equiv \langle a^{\dag}a \rangle)$ corresponds to the boson distribution function $n=\frac{1}{e^{\beta \hbar\omega}-1}$.
At $T=0$, $n=0$, and so $\langle u^2 \rangle = \frac{\hbar}{2M_i\omega}$.
At finite $T$, $\langle u^2 \rangle$ increases with $T$ as
$\langle u^2 \rangle = \frac{\hbar}{2M_i\omega}\coth{\frac{\beta \hbar\omega}{2}}$.
In three dimension, factor 3 needs to be multiplied in the above expressions \cite{SKittel96}.
Note that $\langle u^2 \rangle$ is isotopic-mass dependent through $M_i$ and $\omega$.
But as $T \rightarrow \infty$, $\langle u^2 \rangle$ becomes $\frac{k_B T}{M_i\omega^2}$,
and so it does not have isotopic-mass dependence because of $M_i\omega^2=K$ being constant.

For a real solid, the atomic displacement $u_i$ can be described, as follows \cite{SMahan90}:
\begin{equation}
u_i = i\sum_{k,\lambda}
 \left(\frac{\hbar}{2NM_i\omega_{k\lambda}}\right)^{1/2}
\epsilon_{k\lambda}( a_{k\lambda}+ \a^{\dagger}_{-k\lambda})
e^{i k \cdot R_i},
\label{ui}
\end{equation}
where $\omega_{k\lambda}$ and $\epsilon_{k\lambda}$ are
frequency and polarization vector of ($k,\lambda$) phonon, respectively.
Then $\langle u_i^2 \rangle$ is given by
\begin{eqnarray}
\langle u_i^2 \rangle &=& \sum_{k,\lambda}
 \frac{\hbar}{M_i\omega_{k\lambda}} (n_{k\lambda}+1/2) \nonumber\\
 &=& \sum_{k,\lambda} \frac{\hbar}{2M_i\omega_{k\lambda}} \coth{\frac{\beta \hbar\omega_{k\lambda}}{2}} ,
\label{uiui}
\end{eqnarray}
which is quite similar to Eq. (\ref{ui2-1D})
(here $n_{k,\lambda}=\frac{1}{e^{\beta \hbar\omega_{k\lambda}}-1}$).
Therefore, even in the case of real solid,
$\langle u_i^2 \rangle$ for a specific phonon mode $\omega_{k\lambda}$ of Einstein type
can be described by a simple form of the single harmonic oscillator
$\frac{\hbar}{2M_i\omega_{k\lambda}} \coth{\frac{\beta \hbar\omega_{k\lambda}}{2}}$.



\begin{thebibliography}{99}
\bibitem{Tmd}
The melting temperatures ($T_{m}$'s) for H$_2$O, D$_2$O, and T$_2$O
are 0$~^{\circ}$C, 3.68$~^{\circ}$C, and 4.49$~^{\circ}$C, respectively.
The volume-minimum points corresponding to the density maximum points
($T_{MD}$'s) are 4.0$~^{\circ}$C, 11.19$~^{\circ}$C,
and 13.40$~^{\circ}$C for H$_2$O, D$_2$O, and T$_2$O, respectively.
Molecular volumes at $T_{MD}$'s are 29.9 $\AA^3$, 30.1 $\AA^3$, and	31.1 $\AA^3$, respectively.
\bibitem{Libron}
The other vibrational modes located inbetween the high and low energy phonon modes,
which correspond to the H$_2$O libration and H-O:H bending modes,
are irrelevant to the NTE phenomenon in ice-Ih,
and so are ignored in the present Born-Oppenhemier analysis.

{\cred
}
\bibitem{ice-III} If the intrinsic PTE-anharmonicity of $U_0(V)$ at low $T$ is dominant over
    the NTE-anharmonicity of the QM zero-point energy ($\frac{1}{2}\hbar\omega_s(V)$)
    of high-energy phonon mode, the NTE phenomenon will not emerge.
    Other ice phases under high pressure are likely to be in this situation,
    which is why no NTE phenomena are observed at low $T$.
\end{thebibliography}
\end{document}